\documentclass[slac_one]{revtex4}
\usepackage{graphicx}
\usepackage{epsfig}
\usepackage{fancyhdr}
\usepackage{amssymb}
\newcommand{\pt}{\ensuremath{p_{T}}}
\newcommand{\pttrig}{\ensuremath{p_{T}^{trig}}}
\newcommand{\ptassoc}{\ensuremath{p_{T}^{assoc}}}
\newcommand{\dphi}{\ensuremath{\Delta\phi}}
\newcommand{\deta}{\ensuremath{\Delta\eta}}

\newcommand{\sqrtsNN}{\ensuremath{\sqrt{s_{NN}}}}

\begin{document}
\title{{\small{Hadron Collider Physics Symposium (HCP2008),
Galena, Illinois, USA}}\\ %% Please keep this conference title here
\vspace{12pt}
Jet Modification in Heavy Ion Collisions at RHIC} %% Paper title goes here

\author{M. van Leeuwen}
\affiliation{Universiteit Utrecht, PO Box 80000, Utrecht, Netherlands}

\begin{abstract}
These proceedings present a brief overview of the main results on
jet-modifications in heavy ion collisions at RHIC. In heavy ion
collisions, jets are studied using single hadron spectra and
di-hadron correlations with a high-\pt{} trigger hadrons. At
high \pt, a suppression of the yields due to parton energy loss is
observed. A quantitative confrontation of the data with various
theoretical approaches to energy loss in a dense QCD medium is being
pursued. First results from $\gamma$-jet events, where the photon
balances the initial jet energy, are also presented and compared to
expectations from models based on di-hadron measurements. At
intermediate \pt, two striking modifications of the di-hadron
correlation structure are found in heavy ion collisions: the presence
of a long-range {\it ridge} structure in \deta{}, and a large
broadening of the recoil jet. Both phenomena seem to indicate an
interplay between hard and soft physics.
\end{abstract}

\maketitle

\thispagestyle{fancy}


The goal of research with high-energy nuclear collisions at the
Relativistic Heavy Ion Collider (RHIC) is to study
bulk matter systems where the strong nuclear force as described by
Quantum Chromo Dynamics (QCD) is the dominant interaction. In
particular, it is expected that there is a phase transition of bulk
QCD matter to a deconfined state at high temperature. In heavy ion
collisions, like in other 
hadronic collision, most produced particles are hadrons with low
momenta. Experimental results at RHIC have shown collective behaviour
in the in the soft sector, such as elliptic flow, which is a strong
indication that indeed 'bulk QCD matter' is created in collisions at
RHIC \cite{Adams:2005dq,Adcox:2004mh,Arsene:2004fa,Back:2004je}.

Products of initial state hard scatterings, such as high-\pt{}  hadrons,
photons and heavy mesons, can be used to probe the soft
matter. Initial state production of high-\pt{} partons is relatively
unaffected by the presence of the soft medium, but the partons lose
energy when traversing the medium, dominantly due to gluon
radiation. The goal of high-\pt{} measurements is to study these
interactions and to use them to measure the density and temperature of
the soft matter.

\section{MEDIUM DENSITY FROM HADRON SPECTRA AND DI-HADRONS}
The first indications of parton energy loss in the hot and dense
matter come from inclusive spectra measurement. To quantify the
suppression of particle production in heavy ion collisions, the
nuclear suppression factor 
\begin{equation}
R_{AA}(\pt) = \frac{\left.dN/\pt d\pt dy
  \right|_{Au+Au}}{N_{bin} \left.dN/\pt d\pt dy \right|_{p+p}} ,
\end{equation}
i.e. the ratio of the measured spectrum $dN/\pt d\pt dy$ in Au+Au and
p+p collisions, is used. The factor $N_{bin}$ is the number of
underlying binary nucleon-nucleon collisions in the heavy ion
collisions which is calculated using a geometrical model of the
collision \cite{Miller:2007ri}.

\begin{figure}
\psfig{width=0.5\textwidth,file=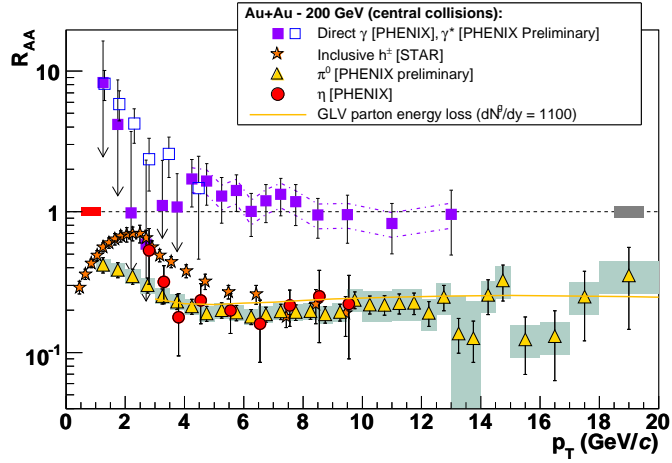}
\caption{\label{fig:gamma_pi}Nuclear suppression ratio $R_{AA}$ (see
  text) as a function of \pt{} for $\gamma$, $\pi^0$ and $\eta$ in
  central Au+Au collisions at \sqrtsNN=200 GeV
  \cite{d'Enterria:2006su}. }
\end{figure}
Figure
\ref{fig:gamma_pi} shows $R_{AA}$ as measured in central Au+Au
collisions at RHIC. For direct photons (solid squares) $R_{AA} \approx 1$,
indicating that the cross section for hard production processes scales
with $N_{bin}$ for particles that do not interact strongly with the medium. For
light hadrons, $\pi^{0}$ and $\eta$, on the other hand, a suppression
by a factor 4--5 ($R_{AA} \approx 0.2$) is seen, signaling strong
interactions of the partons with the medium.

\begin{figure}
\epsfig{width=0.6\textwidth,file=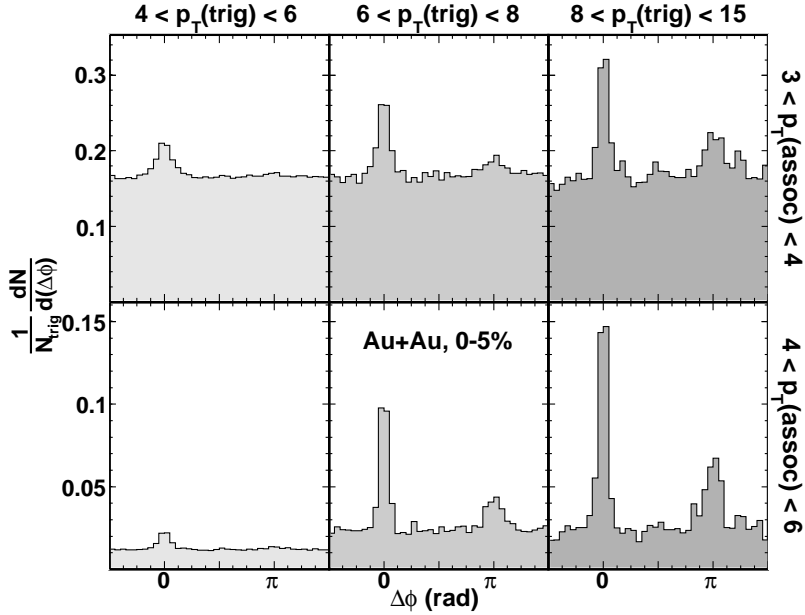}
\caption{\label{fig:dihadr}Per-trigger normalised azimuthal
  distribution of charged particles associated with high-\pt{} trigger
  particles in central Au+Au collisions for different \pt-selections
  for the trigger (\pttrig) and associated particles (\ptassoc)
  \cite{Adams:2006yt}.}
\end{figure}
The jet-structure of high-\pt{} hadron production in heavy ion
collisions is explored using azimuthal di-hadron correlation
measurements. Figure \ref{fig:dihadr} shows distributions of the angle
between a trigger hadron at high \pttrig{} and associated hadrons at
lower \ptassoc. For the higher \pttrig{} selections, there are two
clear correlation peaks due to the near-side ($\dphi \approx 0$) and
the recoil ($\dphi \approx \pi$) jet. The peaks sit on top of a
combinatorial background, which is subtracted to obtain associated
yields.

\begin{figure}
\epsfig{width=0.6\textwidth,file=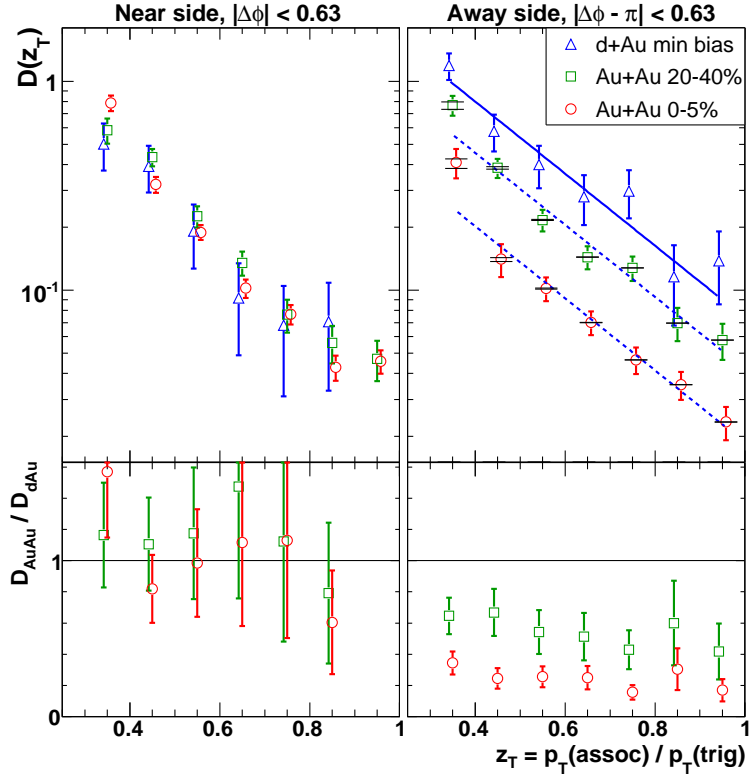}
\caption{\label{fig:assoc}Near-side (left panel) and recoil (right
  panel) associated per-trigger yields for
  trigger hadrons with $8<\pttrig<15$ GeV/$c$, for d+Au collisions and
  Au+Au collisions at two different centralities, 20-40\% and 0-5\%  of
  the total hadronic cross section \cite{Adams:2006yt}. The lower
  panels shows ratios of the Au+Au  results to the d+Au reference.}
\end{figure}
Figure \ref{fig:assoc} shows the near-side (left panel) and recoil
yield (right panel) associated per-trigger yields for trigger hadrons
with $8<\pttrig<15$ GeV/$c$, for d+Au collisions and Au+Au collisions
with two different centralities, 20-40\% and 0-5\% of the total
hadronic cross section. It can clearly be seen that the near side
yield is similar in d+Au and Au+Au collisions, while the recoil yield
shows a suppression which increases with centrality. The per-trigger
normalisation of the yield compensates for the overall scaling of the
number of hard scatterings, due to the larger volume in Au+Au
collisions ( $N_{bin}$ scaling in $R_{AA}$), as well as the leading
hadron suppression. The associated hadron yield on the near-side
measures the secondary fragment distribution given a high-\pt{}
trigger particle. The fact that this distribution does not change from
d+Au to Au+Au collisions, indicates that partons fragment in the same
way in d+Au collisions as in Au+Au collisions after energy loss.  The
recoil yield, on the other hand, measures the distribution of
(leading) hadrons in the jet opposite to the trigger hadron and the
suppression of this yield is a measure of parton energy loss.

\subsection{Quantitative analysis of medium density}
QCD gluon radiation off high-energy partons in a dense QCD medium is
modeled using a number of different approximations. There are four
main calculational frameworks of which two are based on the BDMPS-Z
path integral formalism for multiple scattering in a static medium,
including Landau-Pomeranchuk-Migdal interference effects
\cite{Baier:1996kr,Zakharov:1996fv}. Two approximations exist in this
framework: the multiple-soft-scattering approach, as originally
proposed by BDMPS and further worked out for modeling applications by
by Salgado and Wiedemann \cite{Salgado:2003gb}, and the
few-hard-scattering approach that was proposed originally by Gyulassy,
Levai and Vitev (GLV) \cite{Gyulassy:1999zd}. The two other approaches
are based on a higher-twist formalism, inspired by Deeply Inelastic
Scattering \cite{Wang:2001ifa} and the Hard Thermal Loop
high-temperature QCD formalism (AMY \cite{Arnold:2000dr}). It is
important to realise that there are still a number of open theoretical
questions, such as the effect of collisional energy loss (non-static
medium)\cite{Wicks:2005gt,Djordjevic:2007at} and finite-energy
effects.

\begin{figure}
\psfig{width=0.8\textwidth, file=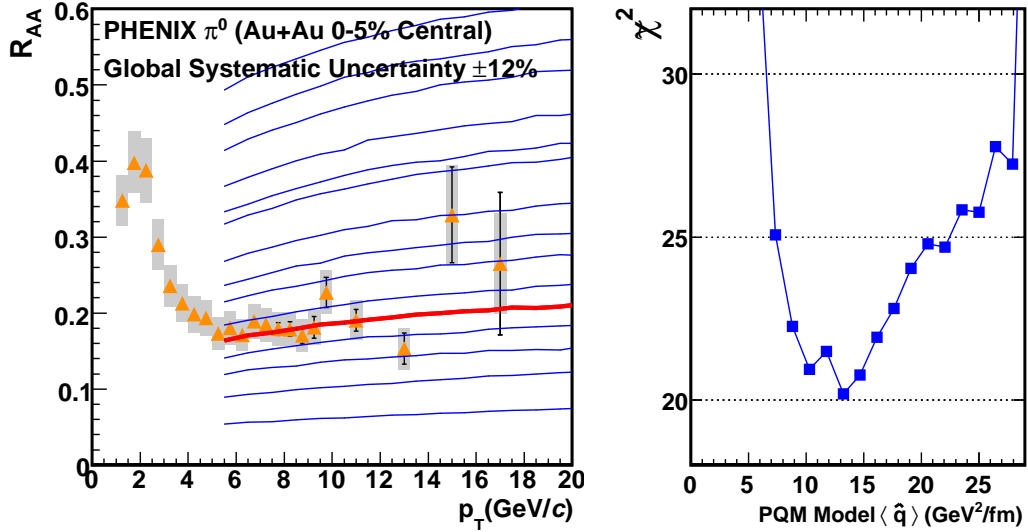}
\caption{\label{fig:RAA_model}Measured nuclear modification factor
  $R_{AA}$ for $\pi^0$, compared to model calculations
  \cite{Dainese:2004te} based on the BDMPS formalism. The right panel
  shows the modified $\chi^2$ of data compared to the model curves,
  including systematic uncertainties \cite{Adare:2008qa}.}
\end{figure}
Now that that the experimental data have a large \pt-reach (up to 18
GeV/$c$ for the $\pi^0$ spectra) and reasonably small
statistical+systematic uncertainties, various groups are exploring
systematic confrontations of theories with the data
\cite{Dainese:2004te,Zhang:2007ja}. A recent advance is the 
quantitative treatment of systematic uncertainties on the measurements
\cite{Adare:2008cg} in these comparisons.

Figure \ref{fig:RAA_model} the expected nuclear modification factor
$R_{AA}$ for several different medium densities from one energy loss
model \cite{Dainese:2004te} compared to the measured values. The right
panel shows the modified $\chi^2$ (which accounts for systematic
effects, see \cite{Adare:2008cg}) as a function of the medium density
which is quoted as the mean transport coefficient $\langle \hat{q}
\rangle$. The transport coefficient $\hat{q}$ is the squared momentum
transfer per unit path length which characterises the energy loss
properties of the medium. The mean energy loss $\Delta E \propto
\alpha_{s} \hat{q} L^2$ for a static medium \cite{Salgado:2003gb}. The
extracted value of the mean transport coefficient $\langle \hat{q}
\rangle = 13.2^{+2.1}_{-3.2}$ GeV$^2$/fm, based on this model. A
number of other models have also been compared to the data
\cite{Adare:2008cg,Nagle:2008fw} and tend to give lower estimates of
the (equivalent) medium density.

\begin{figure}
\psfig{width=0.6\textwidth, file=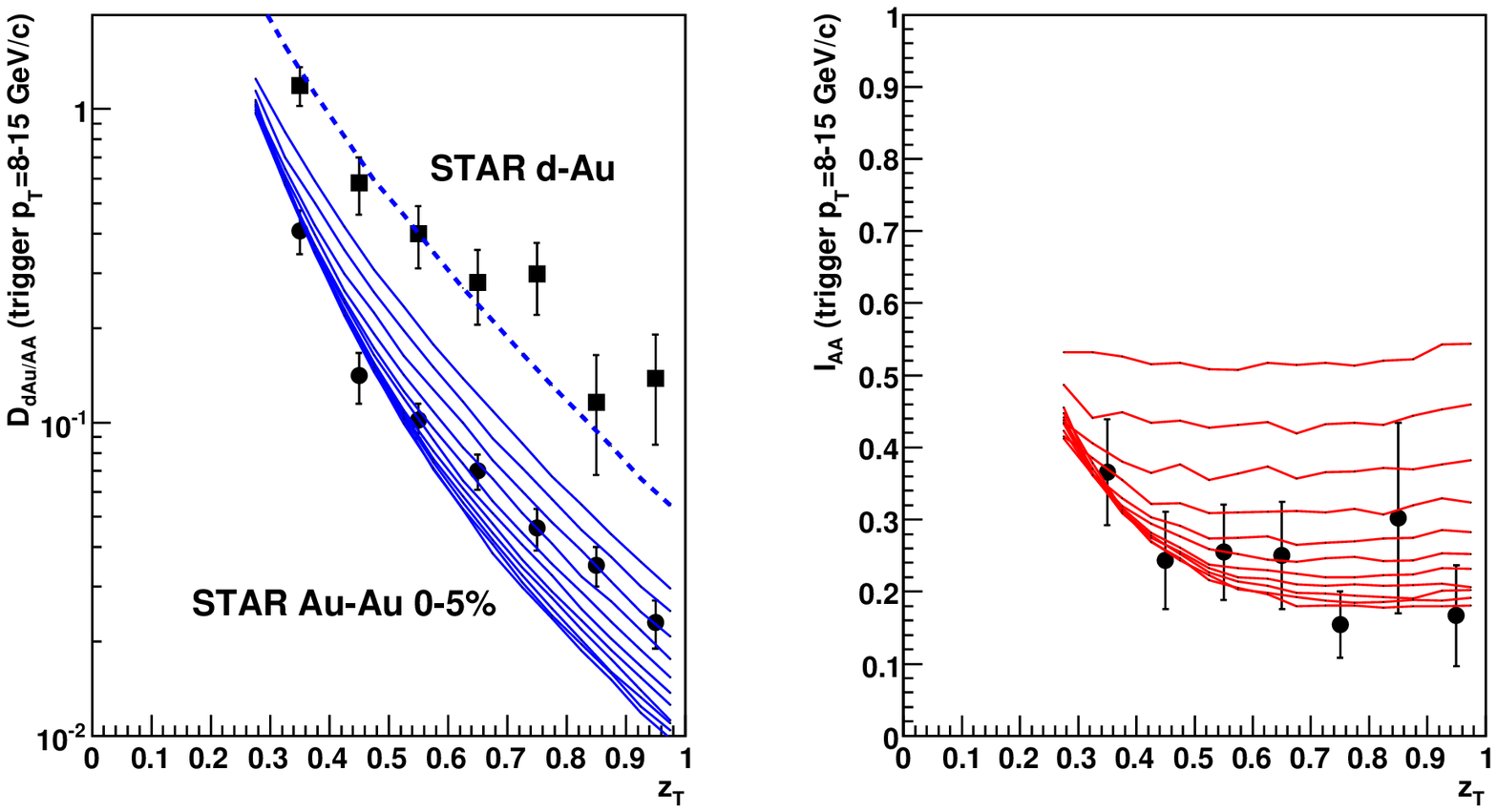}
\psfig{width=0.38\textwidth, file=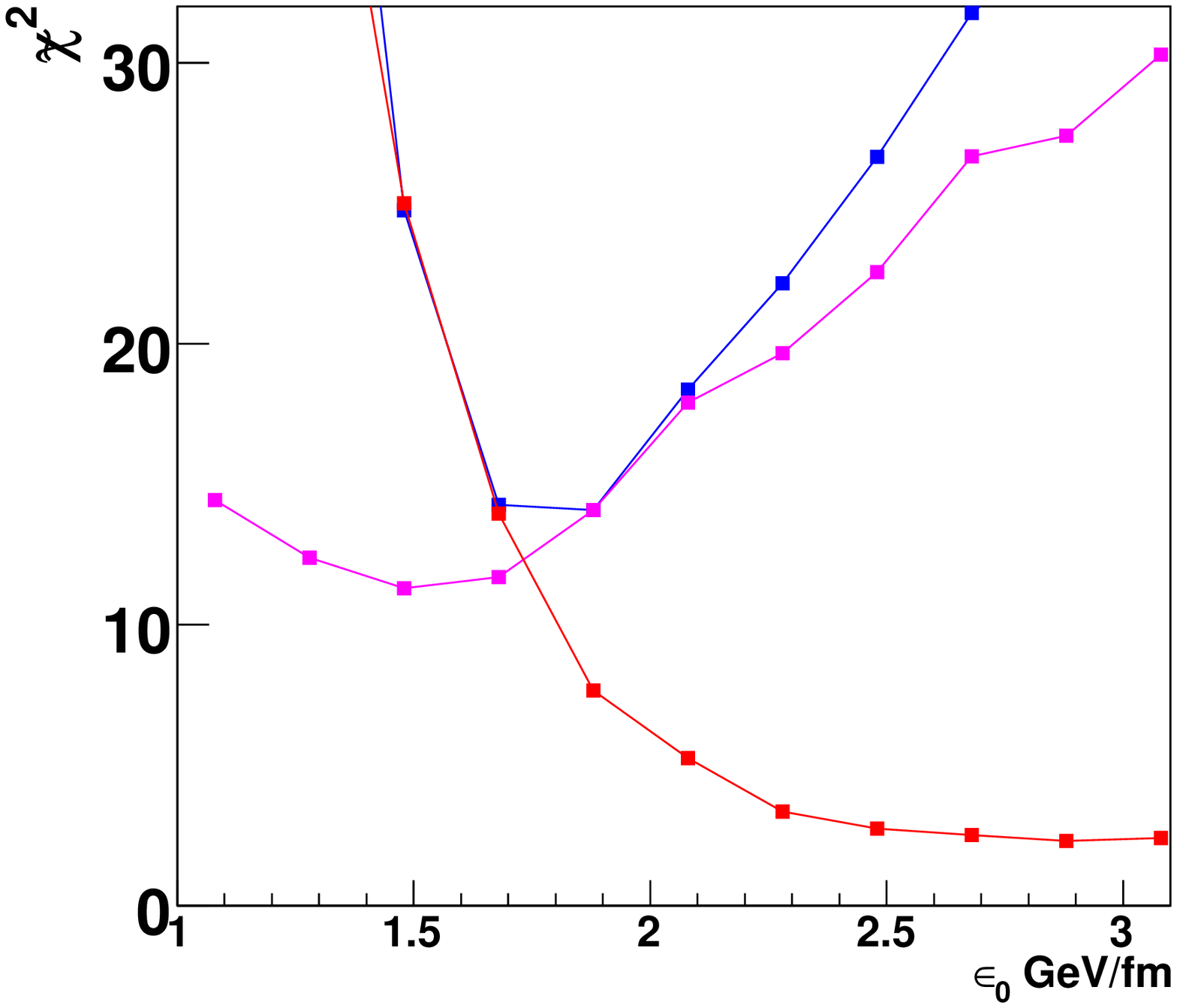}
\caption{\label{fig:IAA_model}Measured suppression of the recoil yield
  with trigger particles $8 < \pttrig <15 $ GeV/$c$ compared to model
  calculations \cite{Zhang:2007ja}. The left panel shows the recoil yield
  $D_{AA}$ for d+Au and Au+Au collisions, the middle panel the
  suppression ratio $I_{AA}$ and the right panel shows the modified
  $\chi^2$ of data compared to the model curves, including systematic
  uncertainties. Three different $\chi^2$ values are given, using the
  recoil spectrum $D_{AA}$ (blue line), including theory uncertainty
  on the reference (magenta line) and using the measured ratio Au+Au
  suppression $I_{AA}$ (red curve) \cite{Nagle:2008fw}. }
\end{figure}
Figure \ref{fig:IAA_model} (left panel) shows the recoil yield for
trigger particles with $8 < \pttrig <15 $ GeV/$c$ in d+Au and Au+Au
collisions compared to model curves with different medium density. The
right panel shows again the modified $\chi^2$ as a function of
density. The model used in this case is a higher twist model, because
the full set of calculations for the recoil yield has so far only been
performed for that model \cite{Zhang:2007ja}. The model parameter in
this case is the typical energy loss $\epsilon_0$. Because the d+Au
reference measurement for this observable has only limited statistical
precision, a few different approaches were taken for the theory
fit. Firstly, one can directly fit the recoil yield, assuming that the
NLO calculation correctly describes the p+p result. The resulting
$\chi^2$ is shown by the blue curve in
Fig. \ref{fig:IAA_model}. The best-fit value for $\epsilon_0 \approx
1.9$ is compatible with the value extracted using the single-hadron
data with this same model $\epsilon_0 = 1.9
^{+0.2}_{-0.5}$ GeV/fm \cite{Nagle:2008fw}. Adding the scale uncertainty on the
calculated d+Au reference yield gives the magenta curve. When
the d+Au measurement is used to calculate the recoil
suppression $I_{AA}$, the red curve is obtained. This last procedure
is the most similar to what was done for the single particle
suppression in Fig. \ref{fig:RAA_model}, but it gives the
weakest constraint on $\epsilon_{0}$. Future high-statistics
measurements of di-hadron correlations in p+p and d+Au collisions at
RHIC will further constrain the theory in this area. 

\section{INTERMEDIATE $P_{T}$: RIDGE AND DOUBLE-HUMP}
Di-hadron correlation measurements at lower
\pt{} show large qualitative differences between p+p reference
measurements and Au+Au results. Two specific striking features
are widely discussed. One effect is the observation that there is
significant associated yield at larger pseudo-rapidity difference
$\Delta\eta\gtrsim0.7$, which is not expected from jet
fragmentation. The other observation is a large broadening of the recoil
distribution at low \pt, to the point where the distribution becomes
doubly-peaked. 

\begin{figure}
\psfig{width=0.45\textwidth, file=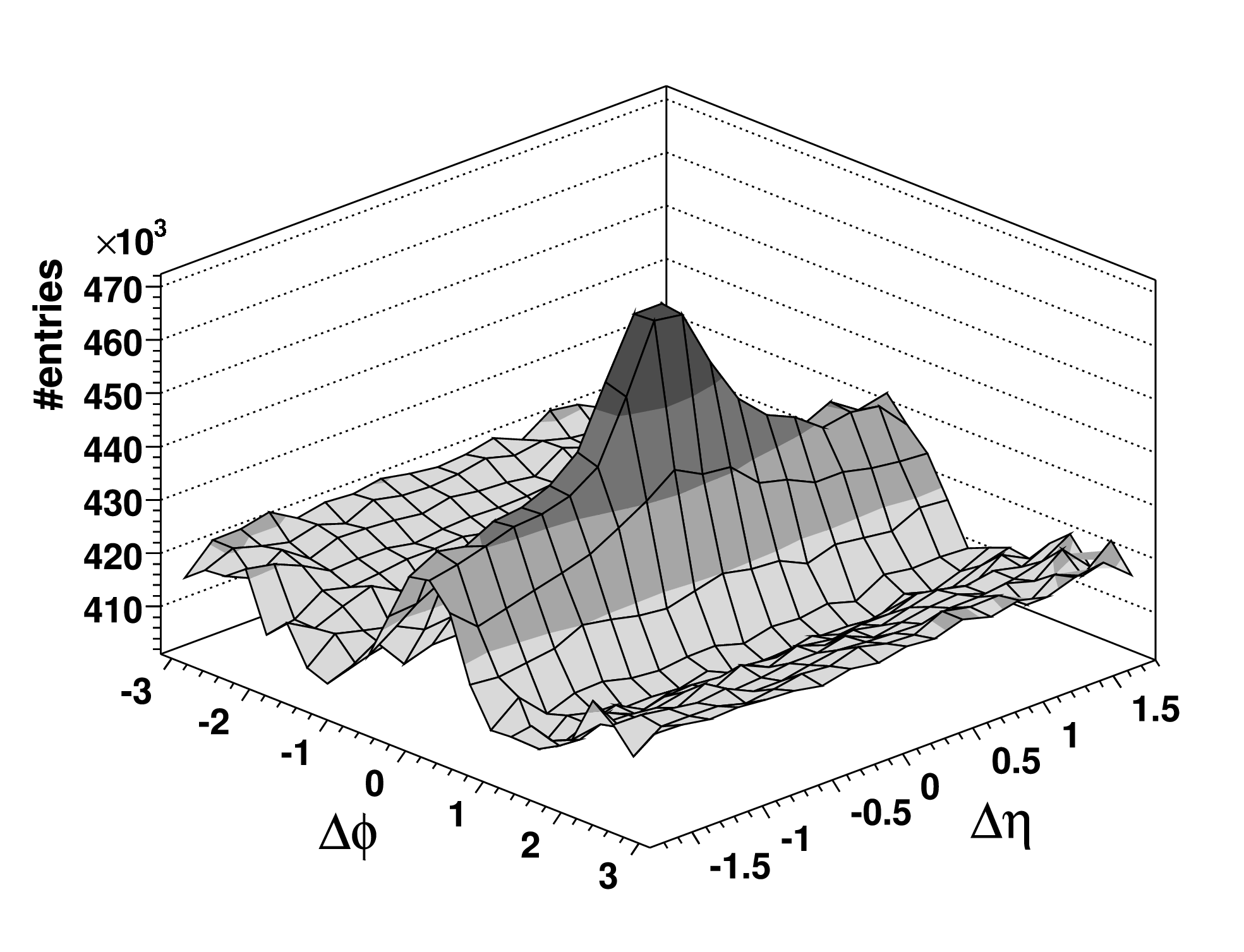}
\caption{\label{fig:ridge}Distribution of associated hadrons
with $2 < \ptassoc < 3$ GeV/$c$ in pseudo-rapidity $\eta$ and azimuthal
angle $\phi$ with respect to a trigger particle with $3<\pttrig<4$
GeV/$c$ in central Au+Au collisions at RHIC \cite{Putschke:2007mi}.}
\end{figure}

Figure \ref{fig:ridge} shows the distribution of associated hadrons
with $2 < \ptassoc < 3$ GeV/$c$ in pseudo-rapidity $\eta$ and azimuthal
angle $\phi$ with respect to a trigger particle with $3<\pttrig<4$
GeV/$c$ in central Au+Au collisions at RHIC \cite{Putschke:2007mi}. At
these \pt, the associated hadrons show not only the jet-like peak
around (\deta,\dphi) = (0,0), but also significant additional
associated yield at larger \deta. The additional yield is
approximately uniformly in \deta{} and therefore it is referred to as
{\it ridge}-yield. The {\it ridge}-effect is unique to heavy ion
collisions and is found to be present for trigger hadrons over the
entire accessible \pt-range (up to ~7 GeV/$c$ at present)
\cite{Putschke:2007mi}. The strength of the jet-like peak increases
with \pttrig, so that the relative contribution of the ridge to the
near-side yield decreases with higher \pttrig. A number of different
possible production mechanisms have been proposed, such as coupling of
radiated gluons to longitudinal flow
\cite{Armesto:2004pt,Majumder:2006wi,Romatschke:2006bb}, medium
heating by the passage of a hard parton combined with longitudinal
flow \cite{Chiu:2005ad} and a radial flow boost to the underlying p+p
event, combined with trigger bias
\cite{Voloshin:2004th,Pruneau:2007ua}. Further experimental work is
going on to distinguish the different scenarios.

\begin{figure}
\psfig{width=0.8\textwidth, file=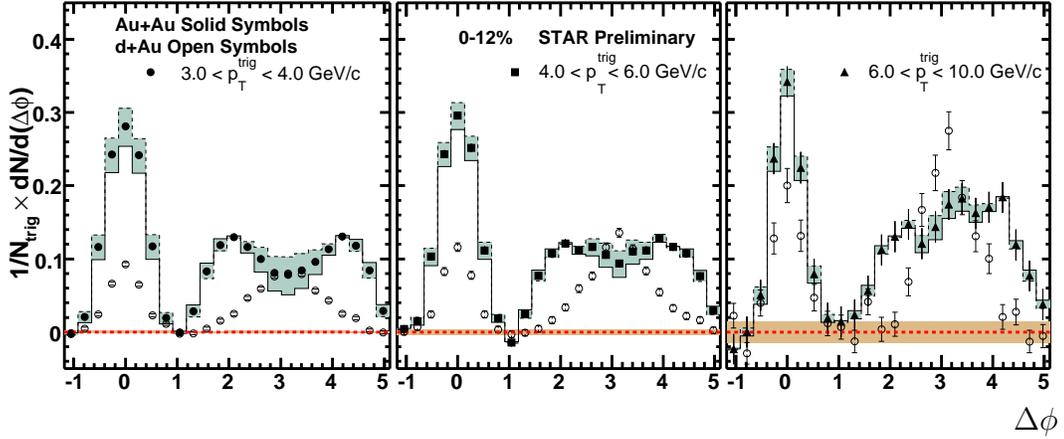, bb=0 25 845 366,clip=}
\put(-15,0){\large\dphi}
\caption{\label{fig:double_peak}Background-subtracted distribution of associated hadrons
with $1 < \ptassoc < 2.5$ GeV/$c$ for three different ranges of
\pttrig, in central Au+Au collisions at \sqrtsNN=200 GeV (full symbols)
and d+Au collisions (open symbols)\cite{Horner:2007gt}. The lines
indicate the uncertainty on the signal shape due to the uncertainty in
the elliptic flow of the background. The shaded bands around 0
indicate the statistical uncertainty on the background level.}
\end{figure}
Another striking finding from di-hadron correlations at intermediate
\pt{} is that the away-side peak is strongly broadened. This is
illustrated in Figure \ref{fig:double_peak}, which shows associated
hadron distributions with $1 < \ptassoc < 2.5$ GeV/$c$ for three
different ranges of \pttrig in central Au+Au collisions
\cite{Horner:2007gt} (full symbols). The open symbols in the Figure
show d+Au results for reference. Clearly, the away-side distribution
is strongly broadened in the Au+Au collisions, compared to d+Au
collisions. For lower \pttrig, there might even be a minimum in the
distribution at $\dphi=\pi$. This observation has lead to the
suggestion that partons propagating through the strongly interaction
medium may give rise to Mach-Cone shock waves
\cite{Stoecker:2004qu,CasalderreySolana:2004qm}. The width of the
away-side distribution would then measure the opening angle and thus
the velocity of sound in the medium. However, it has also been pointed
out that gluon radiation in combination with the kinematic constraint
$\pttrig \approx \ptassoc$ may give rise to a broadened away-side
(or even 'Mercedes-events') as well \cite{Polosa:2006hb}. It is also
important to realise that the raw signal sits on a large background
which is not constant, but has has a $1+\cos(2\dphi)$ distribution due
to elliptic flow. The background has been subtracted in Fig.
\ref{fig:double_peak} and the uncertainty on the extracted signal from
the uncertainty in the strength of elliptic flow is indicated by the
shaded band. However, possible correlations between elliptic flow and
the jet-structure are not taken into account in this estimate.

Three-particle correlation measurements are currently being
developed to further explore the away-side shapes.

%In contrast to what was found at high $\pt \gtrsim 6$ GeV/$c$ where rhe
%di-hadron measurements show a clear di-jet signature and the recoil
%yield is suppressed as expected from theory, 

\section{NEW RESULTS ON ENERGY LOSS AND OUTLOOK: $\gamma$-JET AND JET RECONSTRUCTION}
While the single hadron and di-hadron suppression in Au+Au collisions
can be used to determine the average medium density, more
detailed studies have shown that the both observables, but in
particular the single hadron distribution, are relatively insensitive
to details of the energy loss process, such as the probability
distribution of energy loss \cite{Renk:2007mv}. This is because the
initial jet energy is not constrained in the single-hadron measurement
and only poorly constrained in the di-hadron measurement. In
addition, the typical energy loss (for partons that lose energy; there
is also a finite probability that no radiation occurs) can be as large
as 10 GeV \cite{Renk:2006pk}, which is large compared to the
jet-energies at RHIC.

Currently, two types of measurement are being discussed that provide
access to the initial jet energy before energy loss. The first method
is to use $\gamma$-jet events, where the transverse momentum of the
photon is equal to the initial jet-energy. The drawback of this
method is the small cross section compared to hadron production, which
translates in a limited kinematic reach. An alternative way to measure
the initial parton energy is to perform full jet reconstruction. This
method has potentially large statistics, but suffers from a large
combinatorial background and is therefore mainly discussed for LHC
\cite{Blyth:2006vb,Baur:2000wn}, although work is going on to perform
jet-reconstruction in heavy ion events at RHIC as well.

\begin{figure}
\psfig{file=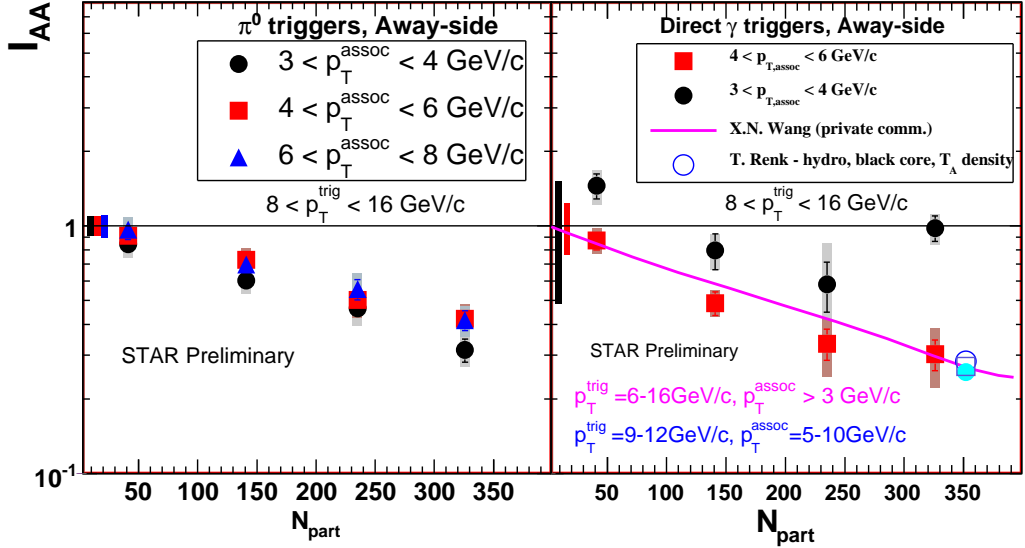,width=0.8\textwidth}
  \caption{\label{fig:iaa_gamma}Centrality dependence of the
  suppression of the recoil yield for leading $\pi^0$ (left panel)
  and direct photons (right panel) with $8 < E_\gamma < 10$ GeV \cite{Hamed:2008yz}.}
\end{figure}
Figure \ref{fig:iaa_gamma} shows first results from $\gamma$-jet pair
in heavy ion collisions, which were presented by the STAR experiment
at the recent Quark Matter conference in Jaipur \cite{Hamed:2008yz}
(see also \cite{Nguyen:2008ic} for a similar analysis from the PHENIX
experiment). The Figure shows the centrality dependence of the
suppression of the recoil yield ($I_{AA}$) opposite $\pi^0$ (left
panel) and direct photons (right panel). The recoil suppression is
found to be similar for $\pi^0$ triggers and direct-photon triggers,
and in agreement with theoretical predictions which are based on the
di-hadron measurements presented in Figure \ref{fig:dihadr}. In the
near future, differential measurements of the suppression as a
function of hadron \pt{} will allow to experimentally constrain the
probability distribution for parton energy loss in the medium
\cite{Renk:2006qg}. 

In summary, high-\pt{} jet production in heavy ion collisions has been
observed at RHIC. The most direct signatures of jet and di-jet
production are provided by di-hadron measurements and more recently in
direct photon-jet events. At high \pt{} a clear suppression of the
recoil jet is observed, indicating that the partons lose energy while
propagating through the dense QCD matter formed in the
collision. Systematic comparisons of single hadron spectra, di-hadron
and $\gamma$-jet measurements to theory calculations are being pursued
to obtain limits on the parton-medium interaction and on the
medium properties themselves. Measurements with reconstructed jets and
more detailed analysis of $\gamma$-jet events, which are expected in
the next few years, are crucial to constrain energy loss models,
because they provide a direct measure of the initial jet energy.

\bibliographystyle{epj}
\bibliography{HCP08_rhic_highpt_mvanleeuwen}

\begin{thebibliography}{38}

\bibitem{Adams:2005dq}
J.~Adams et~al. (STAR), Nucl. Phys. \textbf{A757}, 102 (2005),
  \texttt{nucl-ex/0501009}\relax
\relax
\bibitem{Adcox:2004mh}
K.~Adcox et~al. (PHENIX), Nucl. Phys. \textbf{A757}, 184 (2005),
  \texttt{nucl-ex/0410003}\relax
\relax
\bibitem{Arsene:2004fa}
I.~Arsene et~al. (BRAHMS), Nucl. Phys. \textbf{A757}, 1 (2005),
  \texttt{nucl-ex/0410020}\relax
\relax
\bibitem{Back:2004je}
B.B. Back et~al., Nucl. Phys. \textbf{A757}, 28 (2005),
  \texttt{nucl-ex/0410022}\relax
\relax
\bibitem{Miller:2007ri}
M.L. Miller, K.~Reygers, S.J. Sanders, P.~Steinberg, Ann. Rev. Nucl. Part. Sci.
  \textbf{57}, 205 (2007), \texttt{nucl-ex/0701025}\relax
\relax
\bibitem{d'Enterria:2006su}
D.G. d'Enterria, J. Phys. \textbf{G34}, S53 (2007),
  \texttt{nucl-ex/0611012}\relax
\relax
\bibitem{Adams:2006yt}
J.~Adams et~al. (STAR), Phys. Rev. Lett. \textbf{97}, 162301 (2006),
  \texttt{nucl-ex/0604018}\relax
\relax
\bibitem{Baier:1996kr}
R.~Baier, Y.L. Dokshitzer, A.H. Mueller, S.~Peigne, D.~Schiff, Nucl. Phys.
  \textbf{B483}, 291 (1997), \texttt{hep-ph/9607355}\relax
\relax
\bibitem{Zakharov:1996fv}
B.G. Zakharov, JETP Lett. \textbf{63}, 952 (1996),
  \texttt{hep-ph/9607440}\relax
\relax
\bibitem{Salgado:2003gb}
C.A. Salgado, U.A. Wiedemann, Phys. Rev. \textbf{D68}, 014008 (2003),
  \texttt{hep-ph/0302184}\relax
\relax
\bibitem{Gyulassy:1999zd}
M.~Gyulassy, P.~Levai, I.~Vitev, Nucl. Phys. \textbf{B571}, 197 (2000),
  \texttt{hep-ph/9907461}\relax
\relax
\bibitem{Wang:2001ifa}
X.N. Wang, X.f. Guo, Nucl. Phys. \textbf{A696}, 788 (2001),
  \texttt{hep-ph/0102230}\relax
\relax
\bibitem{Arnold:2000dr}
P.~Arnold, G.D. Moore, L.G. Yaffe, JHEP \textbf{11}, 001 (2000),
  \texttt{hep-ph/0010177}\relax
\relax
\bibitem{Wicks:2005gt}
S.~Wicks, W.~Horowitz, M.~Djordjevic, M.~Gyulassy, Nucl. Phys. \textbf{A784},
  426 (2007), \texttt{nucl-th/0512076}\relax
\relax
\bibitem{Djordjevic:2007at}
M.~Djordjevic, U.~Heinz, Phys. Rev. \textbf{C77}, 024905 (2008),
  \texttt{0705.3439}\relax
\relax
\bibitem{Dainese:2004te}
A.~Dainese, C.~Loizides, G.~Paic, Eur. Phys. J. \textbf{C38}, 461 (2005),
  \texttt{hep-ph/0406201}\relax
\relax
\bibitem{Adare:2008qa}
A.~Adare et~al. (PHENIX) (2008), \texttt{arXiv:0801.4020}\relax
\relax
\bibitem{Zhang:2007ja}
H.~Zhang, J.F. Owens, E.~Wang, X.N. Wang, Phys. Rev. Lett. \textbf{98}, 212301
  (2007), \texttt{nucl-th/0701045}\relax
\relax
\bibitem{Adare:2008cg}
A.~Adare et~al. (PHENIX), Phys. Rev. \textbf{C77}, 064907 (2008),
  \texttt{arXiv:0801.1665}\relax
\relax
\bibitem{Nagle:2008fw}
J.L. Nagle (2008), \texttt{0805.0299}\relax
\relax
\bibitem{Putschke:2007mi}
J.~Putschke, J. Phys. \textbf{G34}, S679 (2007), \texttt{nucl-ex/0701074}\relax
\relax
\bibitem{Armesto:2004pt}
N.~Armesto, C.A. Salgado, U.A. Wiedemann, Phys. Rev. Lett. \textbf{93}, 242301
  (2004), \texttt{hep-ph/0405301}\relax
\relax
\bibitem{Majumder:2006wi}
A.~Majumder, B.~Muller, S.A. Bass, Phys. Rev. Lett. \textbf{99}, 042301 (2007),
  \texttt{hep-ph/0611135}\relax
\relax
\bibitem{Romatschke:2006bb}
P.~Romatschke, Phys. Rev. \textbf{C75}, 014901 (2007),
  \texttt{hep-ph/0607327}\relax
\relax
\bibitem{Chiu:2005ad}
C.B. Chiu, R.C. Hwa, Phys. Rev. \textbf{C72}, 034903 (2005),
  \texttt{nucl-th/0505014}\relax
\relax
\bibitem{Voloshin:2004th}
S.A. Voloshin, Nucl. Phys. \textbf{A749}, 287 (2005),
  \texttt{nucl-th/0410024}\relax
\relax
\bibitem{Pruneau:2007ua}
C.A. Pruneau, S.~Gavin, S.A. Voloshin, Nucl. Phys. \textbf{A802}, 107 (2008),
  \texttt{arXiv:0711.1991}\relax
\relax
\bibitem{Horner:2007gt}
M.J. Horner (STAR), J. Phys. \textbf{G34}, S995 (2007),
  \texttt{nucl-ex/0701069}\relax
\relax
\bibitem{Stoecker:2004qu}
H.~Stoecker, Nucl. Phys. \textbf{A750}, 121 (2005),
  \texttt{nucl-th/0406018}\relax
\relax
\bibitem{CasalderreySolana:2004qm}
J.~Casalderrey-Solana, E.V. Shuryak, D.~Teaney, J. Phys. Conf. Ser.
  \textbf{27}, 22 (2005), \texttt{hep-ph/0411315}\relax
\relax
\bibitem{Polosa:2006hb}
A.D. Polosa, C.A. Salgado, Phys. Rev. \textbf{C75}, 041901 (2007),
  \texttt{hep-ph/0607295}\relax
\relax
\bibitem{Renk:2007mv}
T.~Renk, Phys. Rev. \textbf{C77}, 017901 (2008), \texttt{arXiv:0711.1030}\relax
\relax
\bibitem{Renk:2006pk}
T.~Renk, K.~Eskola, Phys. Rev. \textbf{C75}, 054910 (2007),
  \texttt{hep-ph/0610059}\relax
\relax
\bibitem{Blyth:2006vb}
S.L. Blyth et~al., J. Phys. \textbf{G34}, 271 (2007),
  \texttt{nucl-ex/0609023}\relax
\relax
\bibitem{Baur:2000wn}
G.~Baur et~al., Eur. Phys. J. \textbf{32S2}, 69 (2004)\relax
\relax
\bibitem{Hamed:2008yz}
A.M. Hamed (STAR) (2008), \texttt{arXiv:0806.2190}\relax
\relax
\bibitem{Nguyen:2008ic}
M.~Nguyen (PHENIX) (2008), \texttt{arXiv:0805.1225}\relax
\relax
\bibitem{Renk:2006qg}
T.~Renk, Phys. Rev. \textbf{C74}, 034906 (2006), \texttt{hep-ph/0607166}\relax
\relax
\end{thebibliography}

\end{document}